\documentclass[sigconf]{acmart}
\usepackage{caption}
\usepackage{placeins}
\def\BibTeX{{\rm B\kern-.05em{\sc i\kern-.025em b}\kern-.08emT\kern-.1667em\lower.7ex\hbox{E}\kern-.125emX}}
    
\copyrightyear{2019} 
\acmYear{2019} 
\setcopyright{acmlicensed}
\acmConference[PEARC '19]{Practice and Experience in Advanced Research Computing}{July 28-August 1, 2019}{Chicago, IL, USA}
\acmBooktitle{Practice and Experience in Advanced Research Computing (PEARC '19), July 28-August 1, 2019, Chicago, IL, USA}
\acmPrice{15.00}
\acmDOI{10.1145/3332186.3333161}
\acmISBN{978-1-4503-7227-5/19/07}

\begin{document}

%
% The "title" command has an optional parameter, allowing the author to define a "short title" to be used in page headers.
\title{Research Computing at a Business University}

%
% The "author" command and its associated commands are used to define the authors and their affiliations.
% Of note is the shared affiliation of the first two authors, and the "authornote" and "authornotemark" commands
% used to denote shared contribution to the research.
\author{Jason Wells}
%\authornote{Example authornote}
\email{jwells@bentley.edu}
\affiliation{%
  \institution{Bentley University}
  \department{Academic Technology Center}
  \city{Waltham}
  \state{MA}
  \country{USA}
}
\author{J. Eric Coulter}
%\authornote{Example authornote}
\email{jecoulte@iu.edu}
\affiliation{%
  \institution{Indiana University}
  \department{Science Gateways Research Center}
  \city{Bloomington}
  \state{IN}
  \country{USA}
}

%
% By default, the full list of authors will be used in the page headers. Often, this list is too long, and will overlap
% other information printed in the page headers. This command allows the author to define a more concise list
% of authors' names for this purpose.

%
% The abstract is a short summary of the work to be presented in the article.
\begin{abstract}
Research Computing demands are exploding beyond traditional disciplines due to the proliferation of data in all walks of life.
At Bentley University ("Bentley"), a business university in the Boston area, this expansion has been most readily seen in our
Accounting, Economics, Mathematics, and Natural Sciences departments. The result has been a small effort to build a research
computing capability at this small New England university. This poster will serve as an overview of the steps taken to build
such an effort at a business university, the revelations we have had along the way, and our plans for the future.
\end{abstract}

\begin{CCSXML}
<ccs2012>
<concept>
<concept_id>10003120</concept_id>
<concept_desc>Human-centered computing</concept_desc>
<concept_significance>500</concept_significance>
</concept>
<concept>
<concept_id>10003456.10003457.10003490.10003491.10003494</concept_id>
<concept_desc>Social and professional topics~Systems planning</concept_desc>
<concept_significance>300</concept_significance>
</concept>
<concept>
<concept_id>10003456.10003457.10003490.10003498.10003499</concept_id>
<concept_desc>Social and professional topics~Hardware selection</concept_desc>
<concept_significance>300</concept_significance>
</concept>
<concept>
<concept_id>10003456.10003457.10003490.10003507</concept_id>
<concept_desc>Social and professional topics~System management</concept_desc>
<concept_significance>100</concept_significance>
</concept>
</ccs2012>
\end{CCSXML}

\ccsdesc[500]{Human-centered computing - field studies}
\ccsdesc[300]{Social and professional topics~Systems planning}
\ccsdesc[300]{Social and professional topics~Hardware selection}
\ccsdesc[100]{Social and professional topics~System management}

%
%THESE SHOULD BE BETTER
\keywords{HPC, Long Tail of Science, Non-traditional disciplines, Single Threaded, Windows}

%
% A "teaser" image appears between the author and affiliation information and the body 
% of the document, and typically spans the page. 
%  \Description{Enjoying the baseball game from the third-base seats. Ichiro Suzuki preparing to bat.}

%
% This command processes the author and affiliation and title information and builds
% the first part of the formatted document.
\maketitle

\section{Introduction}
Bentley is a small business university located 30 minutes from Boston in
Waltham, Massachusetts. It has 449 full and part-time
faculty, around 5,000 graduate and undergraduate students, and 24 PhD
candidates. Most of the university's majors are in business
disciplines, or focus on business disciplines from an external perspective
(e.g. Health Economics). As a result, we can say that our Research
Computing effort targets non-traditional High Performance Computing (HPC)
disciplines, and due to a lack of NSF funding, is solidly within the Long
Tail of Science\cite{Heidorn2008}.  

This poster discusses two main questions: whether Bentley researchers have a need for high performance computing,
and whether or not its possible for a school such as Bentley to provide for that need. We will approach the first question through the introduction of a new metric, and the second
from the mindset of providing local hardware resources in a centralized manner. 

\section{Motivation}
Bentley's researchers overwhelmingly conduct their research in three applications (in order): SAS, Stata, and R. The primary issue encountered is that SAS and Stata are limited to a single node ("node bound") and are often running functions or libraries that are limited to a single core ("core bound"). R technically has the same limitations but can overcome them through certain methods. 

It is these limitations that perhaps explain the lack of 'technical efficiency' found with Economics (Bentley's main HPC consumers) by Apon et al. \cite{Apon2015}, and these concerns are not unknown. The Kansas City Federal Reserve Bank's Center for the Advancement of Data and Research in Economics (CADRE) has sought to meet them with nodes equipped with higher speed processors, increased memory amounts, and interactive capabilities \cite{Lougee2018}. Regardless, it is these limitations that drive the research questions above.

\begin{figure}[!ht]
  \centering
  \includegraphics[width=\columnwidth]{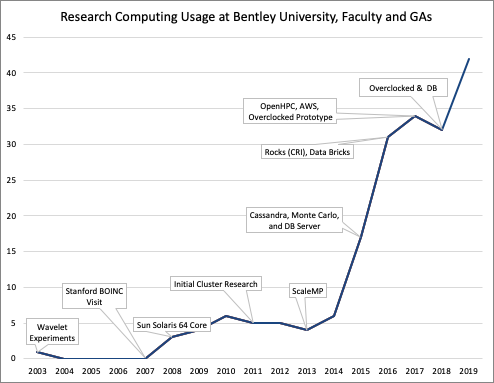}
  \caption{To illustrate our growth, our number of
Faculty and GA/RAs over the years is charted, along with some of the major milestones.}
  \label{fig:faculty}
\end{figure}

\section{The Need}

\begin{figure*}[!ht]
  \centering
  \includegraphics[width=2\columnwidth]{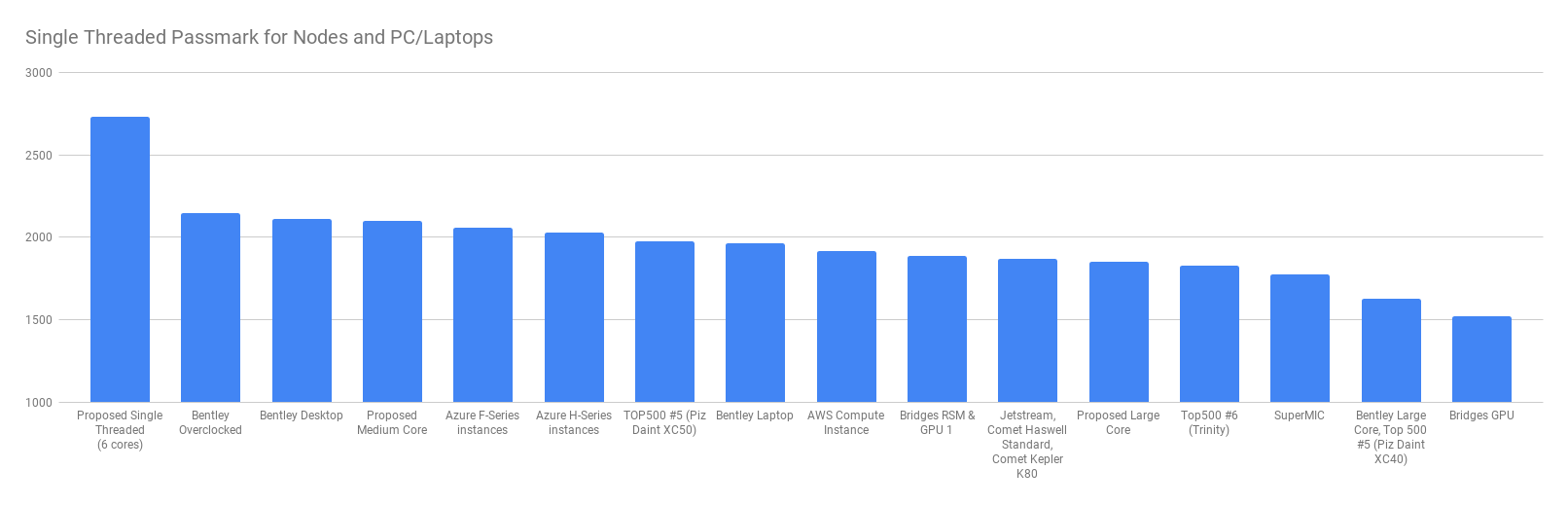}
  \caption{This chart compares performance on the Single Threaded Passmark benchmark between various Bentley resources and
  a variety of supercomputers available to researchers through XSEDE, as well as a few Top500 Intel based supercomputers. For researchers in many disciplines at Bentley,
  the high per-thread performance is more valuable than aggregate performance of thousands of cores.}
  \label{fig:Single_Thread_Passmark}
\end{figure*}

The data sets our faculty use to construct models and simulations have steadily been increasing in
size, causing longer computation times that have pushed the number of possible execution times per year
%changed the following slightly - jec
%changed more, Shauna didn't like it - jrw
down. These execution times can be considered "Research Opportunities Per Year" (ROPY), a useful metric when discussing research-oriented return on investments (ROI) for computational infrastructure. The metric is calculated by the number of units in a year, divided by the programs' execution time in the same units, as so:

\[ ROPY = units_{year}/\textit{execution time}_{unit} \]

To show how the metric is calculated assume one faculty member's (Faculty A) program takes four months to run. Because faculty have access to our HPC resources at all hours of the day, regardless of the day, we can calculate their ROPY as 12 months per year, divided by the four months their program takes to run:

\[ ROPY_{A} = 12_{months}/4_{\textit{months to run program}} \]

Or

\[ ROPY_{A} = 3 \]

Other researchers have much shorter program execution times, such as two hours (Faculty B), giving these researchers ROPYs in
the thousands:

\[ ROPY_{B} = 8760_{\textit{hours in a year}}/2_{hours} = 4380 \]

%Or
%
%\[ ROPY_{faculty_B} = 4380 \]

Our efforts have focused on helping both groups increase their ROPYs. To date, the ROPY 3 researcher is now a ROPY 26, and several researchers with ROPYs in the thousands have reached the tens of thousands.

Given the large increase in the number of researchers at Bentley, as shown in Figure \ref{fig:faculty}, and our success in helping them using the ROPY metric, we have concluded that there is a clear need for HPC resources at Bentley.

\section{Efforts To Date}

\begin{table}[ht]
\caption{The different generations of Bentley's HPC, Data Science, and Storage efforts.}
\label{fig:Table1}
\begin{tabular}{|c|c|c|c|}
\hline
Generation               & HPC / Batch                                                   & \begin{tabular}[c]{@{}c@{}}Data\\ Science\end{tabular} & Storage \\ \hline
First               & ScaleMP                                                   & \begin{tabular}[c]{@{}c@{}}Cassandra\\ Docker\end{tabular}      & HDFS           \\ \hline
Second              & Rocks                                                     & Data Bricks                                                     &                  \\ \hline
Third               & \begin{tabular}[c]{@{}c@{}}OpenHPC\\ Windows\end{tabular} &                                                                 &                  \\ \hline
\end{tabular}
\end{table}

For our first generation of research computing (see Table  \ref{fig:Table1}), we began by convincing our Systems group to give us
its old virtualization Dell M600 hardware in 2013, which was used for HPC and Data Science
purposes. For HPC, we made a three node ScaleMP cluster and for Data Science a 16 node
Cassandra/Docker cluster. The ScaleMP cluster was difficult for our faculty to use however, so for
our second HPC generation in 2016, we opted for a 16 node Rocks cluster. It was at this time that
XSEDE's Campus Resource Integration (CRI)\cite{CB:whatweredoing,XCBC-LessonsLearned} group was brought in to help 
roll out the Rocks cluster, a feat we likely would not have been able to do alone due to the lack of official support channels. Around this time, we also noticed that the Cassandra/Docker cluster was causing too much focus on teaching students how
to use it rather than focusing on the data science they had intended. That resulted in our second Data Science generation, Data Bricks\cite{DataBricks}, a Software as a Service (SaaS) based supplier of Apache Spark, Python, and R-Studio notebooks.

Security concerns forced the retirement of the old Dell M600 hardware in 2018, 
so the remaining
on-campus HPC efforts moved into a third generation with Dell R430s, overclocked
servers from the Quantitative Finance world, and one monster 4U server with 5 NVIDIA Tesla Graphical Processing Unit (GPU) cards
scrounged from around campus. CRI came to the rescue again and helped us build an OpenHPC
\begin{figure*}[!thb]
  \centering
  \includegraphics[width=2\columnwidth]{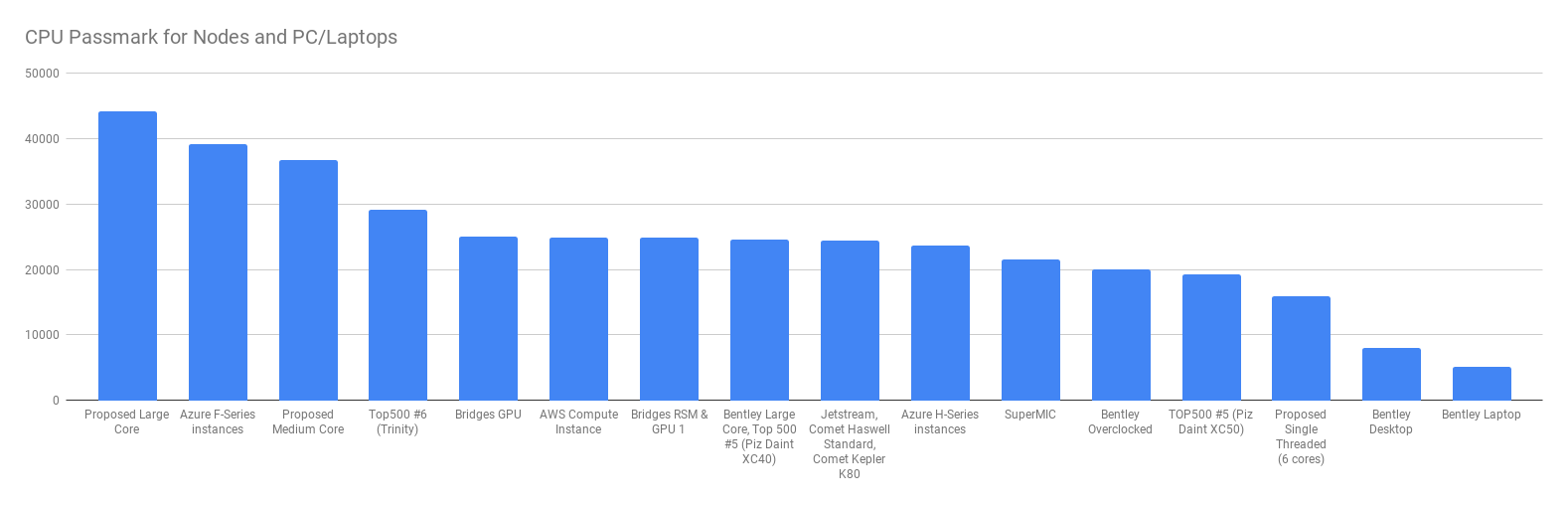}
  \caption{This chart compares performance on the CPU Passmark benchmark between various Bentley resources and
  a variety of supercomputers available to researchers through XSEDE, as well as a few Top500 Intel based supercomputers. For researchers in many disciplines at Bentley,
  the high per-node performance is more valuable than aggregate performance of thousands of cores.}
  \label{fig:CPU_Passmark}
\end{figure*}
cluster with this hardware. Since then, we have added three of the overclocked servers running Windows and delivered to
faculty by Ericom, a Citrix competitor. Additionally, we are adding 36-core count, and database servers to the
OpenHPC and Windows environments. 

First Generation storage efforts started in 2018 with a Hadoop File System (HDFS) instance, but died due to poor performance. This issue is discussed in more detail later in this abstract.

\section{Addressing the Demand} 

Past efforts have largely focused on learning about the advantages and disadvantages of HPC. We have concluded that while clusters make management easy, a greater emphasis is needed on per node and per core performance, as well as on how user friendly our resources are. We address these requirements by focusing on the different elements of a HPC system: Compute, Storage, Networking, and Operating System. A modern concern is whether cloud computing can handle each need, so we briefly address it in each section as well.

\subsection{Compute}

\FloatBarrier

\subsubsection{Methodology}
We utilize a large database of benchmarking figures using the popular Passmark software to compare processors for two benchmarks: 1) the Single Threaded Mark which represents the core bound functions and libraries mentioned above, and 2) a CPU Mark consisting of many CPU tests beyond the Single Threaded Mark, which represents the node bound applications.
\subsubsection{Results}
Bentley's overclocked nodes are aimed at the per core concern, and utilize a single 10-core Intel i7-6950x processor running at 4.394GHz with a single threaded
Passmark score of 2,148 which was amongst the top 10 single threaded Passmark scores in 2017 when purchased. Figure \ref{fig:Single_Thread_Passmark} shows
how this score compares to other supercomputers and even Bentley's standard laptops and desktops. With Bentley desktops being faster on this benchmark than all of the supercomputers it made sense to embrace the overclocked nodes. Recently, however, new CPUs have been released which could negate the need for overclocking. For instance, Intel Xeon E-2186G based nodes (denoted on Figure \ref{fig:Single_Thread_Passmark} as "Proposed Single Threaded") provide a
dramatic increase over even the overclocked nodes.

Another factor to consider is the node based performance. CPU Mark provides the best basis for this comparison.
Figure \ref{fig:CPU_Passmark} shows how our two 18 core Intel E5-2695 v4 node ("Bentley Large Core" on Figures  \ref{fig:Single_Thread_Passmark} and \ref{fig:CPU_Passmark}) compare with the other supercomputer nodes. Our researchers are more likely to favor our servers than apply for an XSEDE allocation, so while our Bentley Large Core node is in the middle of the pack, it suffices for our node bound needs. Here too we see an opportunity to increase our speeds by offering a few dual Intel Xeon Gold 6148 nodes ("Proposed Medium Core" on Figures  \ref{fig:Single_Thread_Passmark} and \ref{fig:CPU_Passmark}), and at least one node with 28 core Intel Xeon Platinum 8180 in quad or eight way configuration ("Proposed Large Core" on Figures  \ref{fig:Single_Thread_Passmark} and \ref{fig:CPU_Passmark}). This server would increase the ROPY of our ROPY 26 client the most, but it will come at a price of USD90K.

We specifically note Azure F-Series, H-Series, and Amazon Web Services (AWS) Compute Instances CPUs on these charts to show how these services aim their HPC solutions. Bentley's single threaded focus is nearly absent on Figure \ref{fig:Single_Thread_Passmark} (our desktops are faster), and while not covered in this poster, the cost for many cores in the highest of CPU Mark cloud instances is very high. Unfortunately those nodes would actually be used for our longest running codes.

\subsection{Storage}
Our first generation storage efforts began with a HDFS instance. It performed poorly chiefly because this parallel virtual file systems speed up access by storing the same file on three drives. We only saw a 3x speedup of the base drive speed, which was paltry for traditional hard drives. We tried to fix this using three RAID sets but ran into another problem. While RAID sets can be used to increase read speeds, the SAS channels they make use of have maximum speeds. As a result, even with 14 SSDs spread across two SAS Raid channels, the fastest read speeds we ever saw was 2.8GBps (1.4GBps per SAS channel), even though the SSDs are each capable of 450MBps individually. With 14 SSDs we should have seen a 13x speed up (5.850GBps). The only traditional solution, as a result, would be many hard drives, in Raid sets, in many servers, a solution a small school like Bentley would not be able to employ. 

\begin{figure}[!th]
  \centering
  \includegraphics[width=\columnwidth]{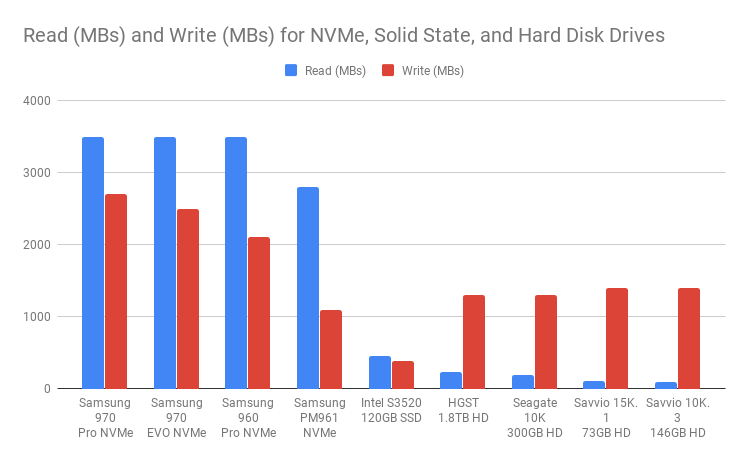}
  \caption{This chart compares read/write performance across a NVMe, SSD, and standard HDD drives. Most large scale resources
  focus more on sheer size of storage rather than speed, though there are a handful of exceptions, such as PSC-Bridges AI-GPU 
  nodes. For clients with moderate data needs and high I/O requirements, NVMe drives offer a huge performance boost.
  }
  \label{fig:drive_speeds}
\end{figure}

But there is a solution. While Solid State Drives are faster than Hard Drives, Bentley desktops and laptops have Non-Volatile Memory Express (NVMe) drives, which are an order of magnitude faster
than even SSDs (Samsung PM961 on Figure \ref{fig:drive_speeds}). This leads to a strange situation where the read speed of the NVMe drives (2800 MBps/2.8GBps/22.4Gbps) in our laptops and desktops is faster than our supercomputer's network fabric (10Gbps). To truly offer a faster storage experience than our client's desktops and laptops, we are considering rolling out converged NVMe drives in a parallel virtual file system (probably BeeGFS), at the same HDFS 3x replication, to yield about 8.4GBps. We can offer this solution using just one server with eight PCIe interfaces, and eight M.2 to PCIe adapters. Please see next section for network implications.

As far as cloud services, we have not yet heard of high speed storage efforts to the degree that we are considering with this service, and past experience with storage costs (10TB is about USD1K/month with AWS EBS and compute for us) give us pause when considering moving to the cloud.

\subsection{Networking}
As a result of the storage speed issue, our 10Gbps network simply will not suffice for future needs. To successfully carry the traffic for NVMe drives (at 3x replication) and metadata, we would need to be running 100Gbps networking speeds to each compute node, which is available in Ethernet or newer Infiniband protocols. Higher replication rates will be possible but the network will limit the usefulness of such rates. Cloud efforts occasionally discuss 25-40Gbps options for Read Direct Memory Applications (RDMA) and networking, but we have not see such speeds for storage connections.

The decision on Ethernet vs. Infiniband is a difficult one. We do not yet have researchers using Message Passing Interface (MPI) libraries or RDMA, so latency has not really been a concern for us. In the end we will likely choose the least expensive total cost for ownership option that gives us the speeds required while also considering future requirements.

\subsection{Operating Systems}
\begin{figure}[!ht]
  \centering
  \includegraphics[width=\columnwidth]{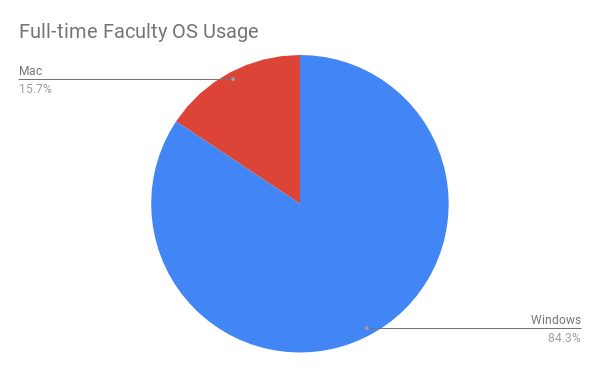}
  \caption{This pie chart documents the number of full-time faculty who currently using Windows, Macs, and Linux as their primary computer Operating System. Please note that Linux is absent from the graph because the number is 0.
  }
  \label{fig:os_graph}
\end{figure}

Bentley researchers mirror their corporate counterparts and utilized Windows nearly entirely until a decade ago, when Apple computers began to creep onto campus. As can be see in Figure \ref{fig:os_graph}, 84.3\% of our full-time faculty use Windows now, with the remainder on Macs. Linux nearly does not exist. This limited the number of faculty willing to work with our non-Windows HPC solutions. For this reason we began offering Windows nodes in 2018 and have seen a dramatic uptake since then (see Figure \ref{fig:faculty}). Most importantly, clients began finding our resources through the Ericom software on their own and simply started using it spontaneously. 

Aside from making introductory training interesting, we also have to deploy hardware solutions in both Linux and Windows as a result. The overclocked servers, Large Core servers, and even GPU resources are offered in both operating systems.

\section{Conclusion}
%A brief conclusion will follow which answers the research questions in the positive and reviews our overall strategy for 
%the future. 
While many of the needs encountered at Bentley are common to non-traditional HPC disciplines, they are not
widely articulated by large research computing centers. Researchers in these areas are often left to compute on desktops, 
pointed towards large systems that are tailored to other needs, or pushed out to the cloud, which, as we have shown, can 
be superseded in speed by well-considered local solutions. Additionally,
moving research onto cloud providers comes with many headaches in the form of billing, security, compliance, systems administration, and 
software/data migration concerns. 
https://www.overleaf.com/4412273942nqvbnjwsrqhg
The biggest gains have been in the realm of compute power. However, upgrading the storage and networking
available to researchers will be invaluable in providing solutions that are truly more powerful than a 
researcher's laptop. It is also clear from our experience that many researchers in non-traditional disciplines are best served by a variety of hardware and operating system options. While progress
in clock speeds has plateaued in recent years, providing access to high Passmark processors of both types is still a viable
model for researcher support in many areas which do not require or strongly benefit from many tightly coupled processors.
Additionally, this model has proven simpler to implement and support than a "traditional" cluster system, and resulted
in happier researchers, faster adoption by researchers, and greatly increased ROPYs. 
Other institutions in the area have also  expressed interest in this model, as they too are discovering that
not all faculty are well supported by cluster computing. In organizations where the primary goal is supporting research,
such systems may be worth investigating in order to broaden the reach of research IT, and to truly bring the computing
revolution to all participants in the research game.

\begin{acks}
This work was partially funded by the Extreme Science and Engineering Discovery Environment (XSEDE), which is supported by 
National Science Foundation grant number ACI-1548562
\end{acks}

\bibliographystyle{unsrt}
\bibliography{my}

\end{document}